\documentclass[aps,pre,twocolumn,superscriptaddress,tight,showpacs]{revtex4-1}
\usepackage{amssymb}
\usepackage{amsfonts}
\usepackage{color}
\usepackage{graphicx}
\usepackage{amsmath}
\usepackage{times}
\usepackage{bm}
\usepackage{float}
\usepackage{lipsum}
\usepackage{import}
\usepackage{enumerate}
\usepackage[colorlinks,linkcolor=blue,anchorcolor=blue,citecolor=blue]{hyperref}

\begin{document}
\title{Reconstructing bifurcation diagrams of chaotic circuits \\with reservoir computing}
\author{Haibo Luo}
\affiliation{School of Physics and Information Technology, Shaanxi Normal University, Xi'an 710062, China}
\author{Yao Du}
\affiliation{School of Physics and Information Technology, Shaanxi Normal University, Xi'an 710062, China}
\author{Huawei Fan}
\affiliation{School of Science, Xi'an University of Posts and Telecommunications, Xi'an 710121, China}
\author{Xuan Wang}
\affiliation{School of Physics and Information Technology, Shaanxi Normal University, Xi'an 710062, China}
\author{Jianzhong Guo}
\affiliation{School of Physics and Information Technology, Shaanxi Normal University, Xi'an 710062, China}
\author{Xingang Wang}
\email{E-mail address: wangxg@snnu.edu.cn}
\affiliation{School of Physics and Information Technology, Shaanxi Normal University, Xi'an 710062, China}

\begin{abstract}
Model-free reconstruction of the bifurcation diagrams of Chua's circuits by the technique of parameter-aware reservoir computing is investigated. We demonstrate that: (1) reservoir computer can be utilized as a noise filter to recover the system trajectory from noisy signals; (2) for a single Chua circuit, the machine trained by the noisy time series measured at several sampling states is capable of reconstructing the whole bifurcation diagram of the circuit with a high precision; (3) for two coupled chaotic Chua circuits of mismatched parameters, the machine trained by the noisy time series measured at several coupling strengths is able to anticipate the variation of the synchronization degree of the coupled circuits with respect to the coupling strength over a wide range. The studies verify the capability of the technique of parameter-aware reservoir computing in learning the dynamics of chaotic circuits from noisy signals, signifying the potential application of this technique in reconstructing the bifurcation diagram of real-world chaotic systems.
\end{abstract}
\date{\today}
\maketitle

\section{Introduction}

In exploring chaotic systems, one of the central tasks is to characterize how the system dynamics is varying with the system parameters, namely finding the bifurcation diagram of the system dynamics~\cite{Ott:Book,MS:Book}. The study of the bifurcation diagram is not only of theoretical interest as it reveals the route from regular behaviors to chaos, but also of practical significance as it pinpoints the tipping points where a small change in the system parameters might result in a drastic change in the system dynamics~\cite{TP:MG2006,TP:Review2009}. The latter is of particular concern to our modern society, as accumulating evidence indicates that many real-world complex systems are already in the vicinity of their tipping points, e.g., the global climate~\cite{TP:Climate1,TP:Climate2}, complex ecological systems~\cite{TP:Eco1,TP:Eco2}, and financial markets~\cite{TP:FinancialSys1,TP:FinancialSys2}. When the exact equations governing the system dynamics are known, the bifurcation diagram can be constructed by the approach of model simulations. Yet in realistic situations the exact equations of the system dynamics are generally unknown, and what is available are only measured data. Different from model-based studies in which the signals are noise-free and the system parameters can be tuned arbitrarily according to the research request, signals measured from realistic systems are inevitably contaminated by noise. In addition, due to the cost of data acquisition and practical restrictions, it is infeasible to construct the bifurcation diagram of a realistic system by a fine scan of the system parameters over a wide range. These practical concerns make model-free reconstruction of the bifurcation diagram of realistic chaotic systems a challenging question of active research in the field of nonlinear science and complex systems~\cite{BD:FD1980,BD:NHP1980,BD:RT1994,BD:EB1999,BD:GL2004,BD:RC2019,RC:Follmann2019,BD:YI2020,RC:ZHScienceChina2021,KLW2021,RC:Kim2021,HWFan2021,RC:ZH2021,Roy2022}.            
   
To reconstruct the bifurcation diagram of chaotic systems based on measured data, one approach is to rebuild the model first, including inferring the terms contained in the dynamical equations and estimating the system parameters, and then reconstructing the bifurcation diagram through the approach of model simulations~\cite{Reverse:SLB2016,CS:WW2016,GHU:2018}. The advantage of this model-rebuilding approach is that the equations governing the system dynamics can be obtained explicitly, while the drawbacks are that the data should be of high quality (with weak noise) and some prior knowledge of the system dynamics should be available, e.g., the form of the nonlinear terms in the equations. An alternative approach to reconstructing the bifurcation diagram is exploiting the machine learning techniques~\cite{BD:GL2004,BD:RC2019,RC:Follmann2019,BD:YI2020,RC:ZHScienceChina2021,KLW2021,RC:Kim2021,HWFan2021,RC:ZH2021,Roy2022}. Owning to the superpower of regression analysis, machine learning techniques are able to infer from data not only the dynamics of the chaotic systems but also the system parameters, and therefore are capable of reconstructing the bifurcation diagrams. Compared to the model-rebuilding approach, the advantages of the machine-learning approach are that no prior knowledge of the system dynamics is required and the techniques can be applied to noisy signals in general, yet the disadvantages are that the system dynamics are unknown, i.e., the machines are working as ``black boxes", and a large amount of data are normally required for training the machines. 

Reservoir computing (RC)~\cite{RC:Maass2002,RC:Jaeger}, a special technique based on recurrent neural networks in machine learning, has been exploited recently for predicting chaos and reconstructing the bifurcation diagram of chaotic systems~\cite{RC:Lu2017,RC:Pathak2017,RC:Pathak2018,RC:SynSmall2019,RC:Fan2020,KLW2021,RC:Kim2021,HWFan2021,RC:ZH2021,Roy2022}. From the point of view of dynamical systems, a reservoir computer can be regarded as a complex network of coupled nonlinear units, which, driven by the input signals, generates the outputs through a readout function~\cite{RC:Book}. Compared to other types of deep learning techniques such as convolutional neural networks (CNNs), RC contains only a single hidden layer, namely the reservoir. Except for the output matrix which is to be estimated from the data through a training process, the machine is fixed at the construction, including the input matrix, the reservoir network, and the updating rules. Though structurally simple, RC has shown its great potential in many data-oriented applications~\cite{RC:Book}, e.g., speech recognization, channel equalization, robot control, and chaos prediction. In particular, it has been shown that a properly trained RC is able to predict accurately the state evolution of typical chaotic systems for about half a dozen Lyapunov times~\cite{RC:Jaeger,RC:Pathak2018}, which is much longer than the prediction horizon of the traditional methods developed in nonlinear science. Besides predicting the short-term state evolution, RC is also able to replicate faithfully the long-term statistical properties of chaotic systems, e.g., the dimension of strange attractors and the Lyapunov exponents~\cite{RC:Pathak2017}. This ability, known as climate replication, has been exploited very recently to predict the critical transitions and bifurcation points in complex dynamical systems~\cite{KLW2021,RC:Kim2021,HWFan2021,Roy2022}. In particular, by incorporating a parameter-control channel into the standard RC, it has been demonstrated that the machine trained by the time series of several sampling states of a chaotic system is able to infer the dynamical properties of the other states not included in the training set. This new technique, which is named parameter-aware RC (PARC) in Ref.~\cite{KLW2021}, has been applied successfully to predict the critical transition of system collapses, infer the bifurcation diagram of chaotic systems~\cite{RC:Kim2021,Roy2022}, and anticipate the critical coupling for synchronization in coupled oscillators~\cite{HWFan2021}. Whereas the efficacy of the PARC technique has been well demonstrated in these applications, the studies are restricted to modeling systems of noise-free signals and exact parameters. As noise perturbations and parameter uncertainty are inevitable in realistic systems, a question of general interest therefore is whether the PARC technique can be applied to realistic chaotic systems.  

It is worth noting that the impact of noise on the performance of RC in predicting chaotic systems is twofold. On the one hand, noise-corrupted signals blur the system trajectories, making it difficult to infer accurately the equations of the system dynamics~\cite{NoiseFilter:2020,NoisePDC:2021,Noise:Nathe2023,Noised:Lin2022}. A typical case of this kind is measurement noise, which is commonly regarded as destructive to machine learning. To cope with measurement noise, techniques such as low-pass filters are usually adapted to process the data before feeding them into the machine~\cite{NoiseFilter:2020,Noise:Nathe2023}. On the other hand, noise might play a constructive role in machine learning in some circumstances~\cite{Noise:CB1995,Noise:AR2021,HWFan2022,Noise:SR2023}. For measurement noise, studies have shown that in the training phase the role of noise is similar to that of Tikhonov regularization~\cite{Noise:CB1995}, and the performance of the machine reaches its maximum at the moderate noise~\cite{Noise:SR2023}. For dynamical (intrinsic) noise, studies have shown that the introduction of a certain amount of noise is helpful for exploring the global information of the system dynamics, and therefore is beneficial for machine learning, e.g., extending the transient dynamics and inferring the ``unseen" attractors~\cite{Noise:AR2021,HWFan2022}. The nontrivial relationship between noise and machine learning makes the inference of chaotic dynamics from noisy signals not only a practical concern in applications, but also an effective approach for exploring the working mechanism of the machines. For that, growing attention has been paid in recent years to the prediction and inference of chaos based on noisy signals~\cite{NoiseFilter:2020,NoisePDC:2021,Noise:Nathe2023,Noised:Lin2022,Noise:AR2021,HWFan2022,Noise:SR2023}. The studies, however, are mostly conducted for modeling systems with artificial noise, with the validity of the results in realistic system is yet to be checked.      

In our present work, employing the classic Chua circuits as examples, we attempt to reconstruct from measured data the bifurcation diagrams of the circuits by the PARC technique proposed recently in machine learning. Two specific scenarios are considered and investigated. In the first scenario, we collect the time series from a single circuit under several sampling parameters, and the mission is to reconstruct the whole bifurcation diagram in the parameter space. In the second scenario, we collect the time series of two coupled chaotic circuits under several coupling parameters, and the mission is to anticipate the variation of the synchronization degree of the coupled circuits with respect to the coupling parameter over a large range. We are going to demonstrate that, despite the presence of noise (measurement and dynamical noise) and parameter mismatch (between two coupled circuits), the PARC technique is capable of reconstructing the bifurcation diagrams with high precision in both scenarios. The rest of the paper is organized as follows. In the following section, we will describe the experimental setups and the way how the data are acquired. The technique of PARC will be introduced briefly in Sec. III. Our main results on the application of the PARC technique will be presented in Sec. IV, including the filtering effect of RC on the noisy signals, the reconstruction of the bifurcation diagram for a single circuit, and the inference of the synchronization relationship between two coupled chaotic circuits. Finally, concluding remarks will be given in Sec. V. 

\section{Experimental setups}

\begin{figure}[tbp]
\begin{center}
\includegraphics[width=\linewidth]{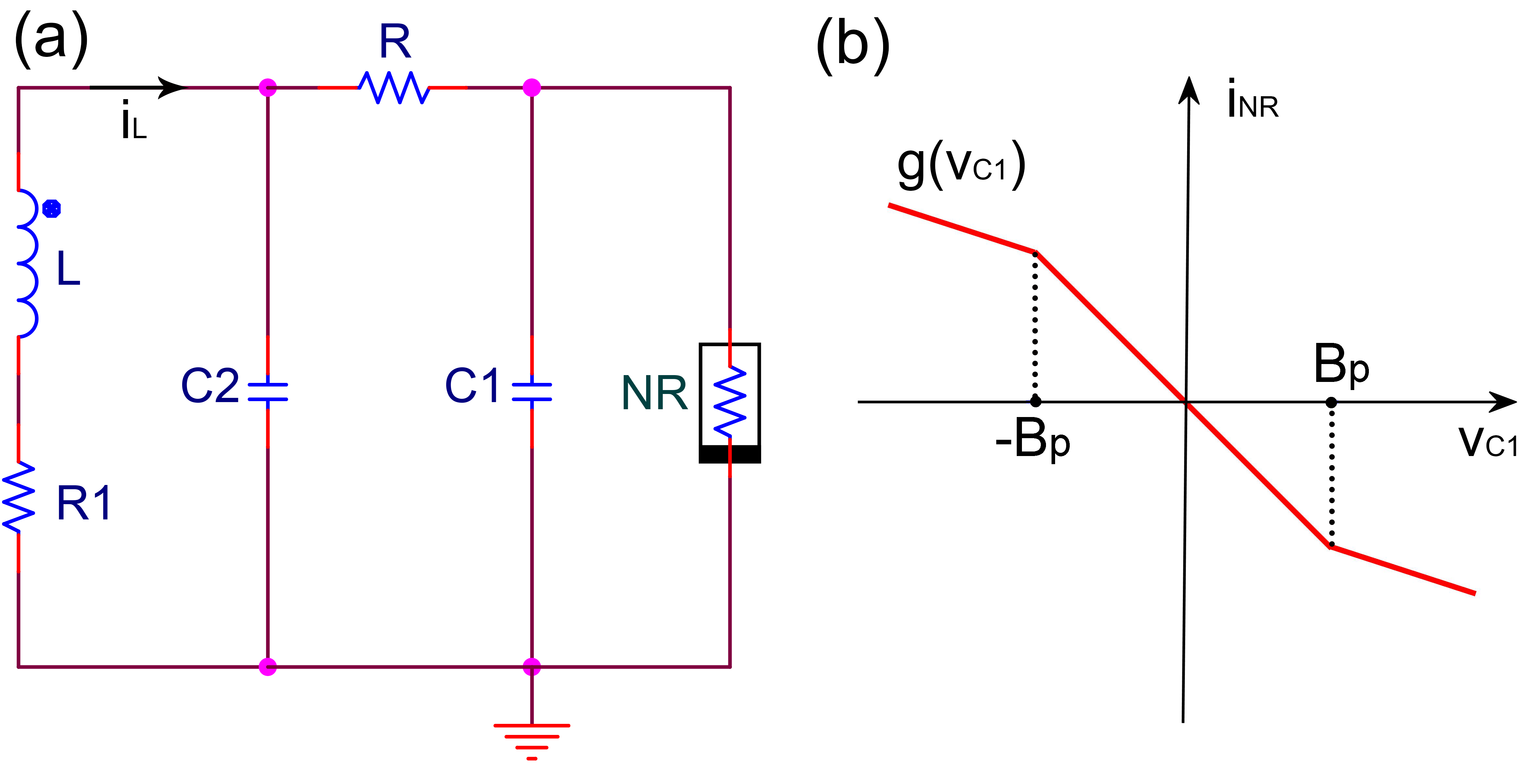}
\centering\caption{(a) Schematic of Chua's circuit. NR denotes the nonlinear resistor. The linear resistor $R$ plays the role of the bifurcation parameter, which is adjusted to generate different dynamics. (b) The piecewise-linear characteristic curve of the NR.} 
\label{fig1}
\end{center}
\end{figure}

The Chua's circuit adopted in our studies is schematically shown in Fig.~\ref{fig1}(a), which consists of two capacitors ($C_{1}$ and $C_{2}$), two linear resistors ($R$ and $R_1$), one inductor ($L$), and a nonlinear resistor (NR)~\cite{ChuaModel,ChuaExp1,ChuaExp2,ChuaExp3}. The equations of the system dynamics read
\begin{equation}
\begin{cases}
C_1\dfrac{dv_{C_1}}{dt}=\dfrac{1}{R}(v_{C_2}-v_{C_1})-g(v_{C_1}),\\
C_2\dfrac{dv_{C_2}}{dt}=\dfrac{1}{R}(v_{C_1}-v_{C_2})+i_L,\\
L\dfrac{di_L}{dt}=-v_{C_2}-R_1i_L,
\end{cases}
\label{model}
\end{equation} 
with $g(v_{C_1})=m_0v_{C_1}+0.5(m_1-m_0)(|v_{C_1}+B_p|-|v_{C_1}-B_p|)$ the characteristic curve of the nonlinear resistor. The characteristic curve of the nonlinear resistor is schematically plotted in Fig.~\ref{fig1}(b), in which the parameters are $m_0=-0.41\,\mbox{mS (mA/V)}\pm 10\%$, $m_1=-0.76\,\mbox{mS}\pm 10\%$, and $ B_p=1.7\,\mbox{V}\pm 5\% $. In our experiments, we fix the components $R_1=10\,\Omega\pm1\%$, $C_1=10\,\mbox{nF}\pm5\%$, $C_2=100\,\mbox{nF}\pm5\%$, $L=20\,\mbox{mH}\pm10\%$, while changing $R$ over the range $(1.73\,\mbox{k}\Omega, 1.77\,\mbox{k}\Omega)$ to generate different dynamics. The variables measured in the experiments are $v_{C_1}$ (the voltage of capacitor $C_1$), $v_{C_2}$ (the voltage of capacitor $C_2$), and $v_{R_1}=i_LR_1$ (the voltage of resistor $R_1$), which are acquired by the sampling frequency $f_0=50\,\mbox{kHz}$. For each value of $R$, we first let the circuit operate for a transient period of $1000\,\mbox{ms}$, and then record the system state, $(v_{C_1}, v_{C_2}, v_{R_1})$, for a period of $100\,\mbox{ms}$. As such, each time series contains $n=5000$ data points.

\begin{figure}[tbp]
\begin{center}
\includegraphics[width=0.75\linewidth]{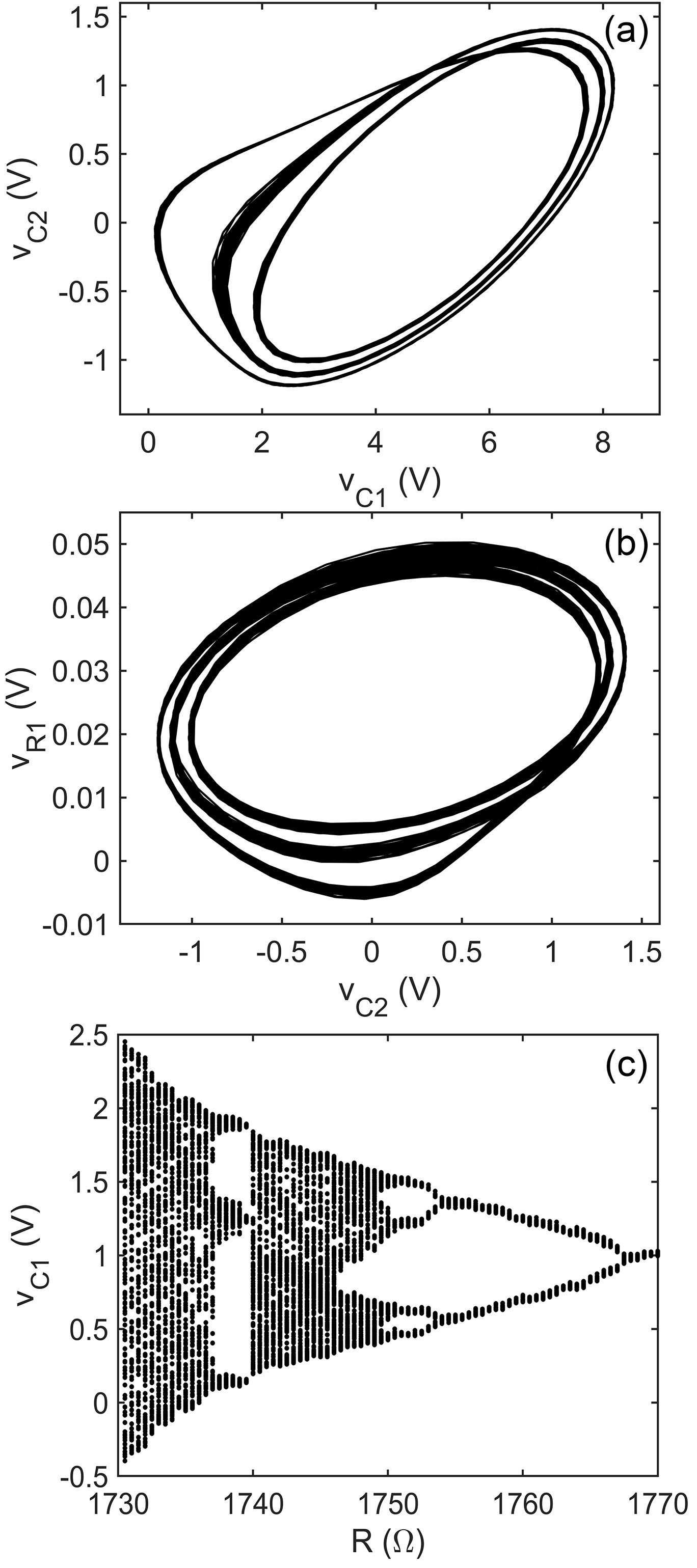}
\centering\caption{Setting $R=1.738\,\mbox{k}\Omega$ in Chua's circuit, the system trajectories plotted on the planes $(v_{C_1}, v_{C_2})$ (a) and $(v_{C_2}, v_{R_1})$ (b). (c) By the data measured from experiments, the bifurcation diagram of Chua's circuit plotted according to the local minimums of $v_{C_1}$.} 
\vspace{-0.2cm}
\label{fig2}
\end{center}
\end{figure}

Setting $R=1.738\,\mbox{k}\Omega$ in the circuit, we plot in Figs.~\ref{fig2}(a) and (b) the system trajectories projected onto the 2D phase spaces $(v_{C_1}, v_{C_2})$ and $(v_{C_2}, v_{R_1})$, respectively. We see that the trajectories are blurred by noise severely, rendering it difficult to figure out accurately the periodicity of the trajectories. (The trajectories seem to be period-3, but might be period-6 or weakly chaotic.) We also see from the figures that compared to the variables $v_{C_1}$ and $v_{C_2}$, the variable $v_{R_1}$ is more corrupted by noise. For this reason, we choose the variable $v_{C_1}$ to investigate experimentally the bifurcation diagram. Decreasing $R$ from $1.77\,\mbox{k}\Omega$ to $1.73\,\mbox{k}\Omega$ by the decrement $\Delta R=0.5\,\Omega$, we measure the time series of $v_{C_1}$ for each value of $R$ and, by recording the local minimums of $v_{C_1}$, plot in Fig.~\ref{fig2}(c) the bifurcation diagram of the circuit. We see that, while the figure shows roughly the route from limit cycle to chaos through the period-doubling bifurcations, the bifurcation details are not clearly shown. For instance, we can not infer from the figure when will the system dynamics present the period-8 orbit and what happens in the window $R\in [1735\,\Omega,1741\,\Omega]$. {\it The first objective of our present work is to reconstruct the bifurcation diagram of Chua's circuit with a high quality (precision), based on the noisy series acquired at several values of $R$ in experiments.}

\begin{figure}[tbp]
\begin{center}
\includegraphics[width=0.95\linewidth]{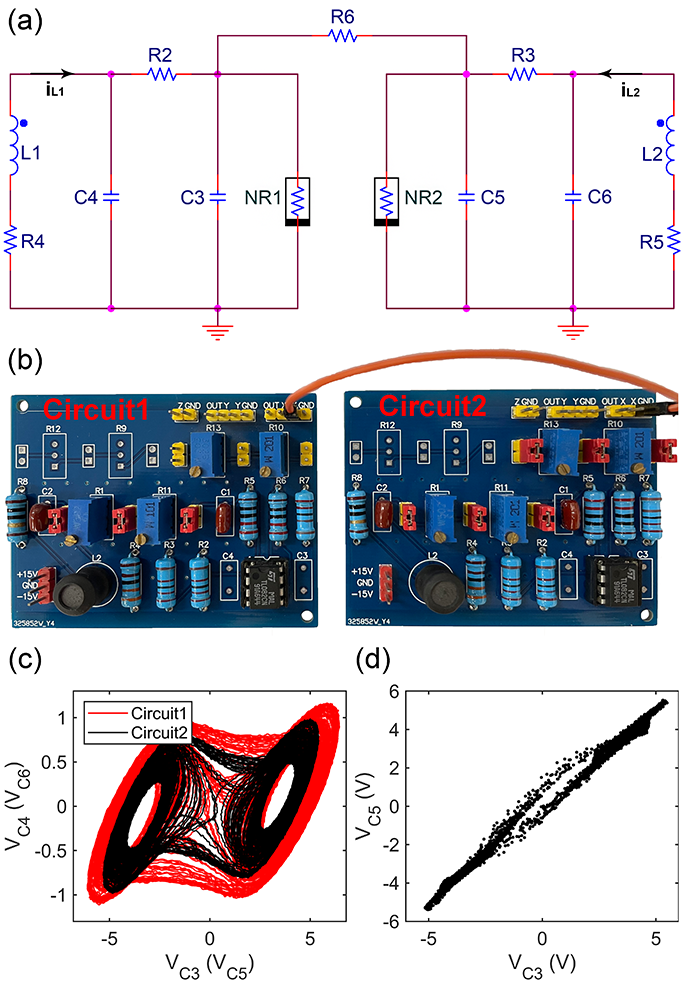}
\centering\caption{(a) Schematic of two coupled Chua circuits. (b) The experimental setup. (c) The trajectories of isolated chaotic circuits on the 2D phase spaces $(v_{C_3}, v_{C_4})$ and $(v_{C_5}, v_{C_6})$. (d) Setting $R_6=10.2\,\mbox{k}\Omega$ in the experiment, $v_{C_3}$ versus $v_{C_5}$ plotted according to the measured data.} 
\vspace{-0.8cm}
\label{fig3}
\end{center}
\end{figure}

The second experiment we conduct is the synchronization of two coupled chaotic Chua circuits. The diagram of the coupled circuits is schematically shown in Fig.~\ref{fig3}(a), and a photo of the experimental setup is given in Fig.~\ref{fig3}(b). The dynamics of the coupled circuits are governed by the equations
\begin{equation}
\begin{cases}
C_3\dfrac{dv_{C_3}}{dt}=\dfrac{1}{R_2}(v_{C_3}-v_{C_4})-g(v_{C_3})+\dfrac{1}{R_6}(v_{C_5}-v_{C_3}),\\
C_4\dfrac{dv_{C_4}}{dt}=\dfrac{1}{R_2}(v_{C_4}-v_{C_3}+i_{L_1}),\\
L_1\dfrac{di_{L_1}}{dt}=-v_{C_4}-R_4i_{L_1},\\
C_5\dfrac{dv_{C_5}}{dt}=\dfrac{1}{R_3}(v_{C_5}-v_{C_6})-g(v_{C_5})+\dfrac{1}{R_6}(v_{C_3}-v_{C_5}),\\
C_6\dfrac{dv_{C_6}}{dt}=\dfrac{1}{R_3}(v_{C_6}-v_{C_5}+i_{L_2}),\\
L_2\dfrac{di_{L_2}}{dt}=-v_{C_6}-R_5i_{L_2},
\end{cases}
\label{model2}
\end{equation}  
with $g(v_C)$ the piecewise-linear function characterizing the nonlinear resistors. [The parameters of the nonlinear resistors are identical to the one used Fig.~\ref{fig1}(b)]. Here, to better demonstrate the synchronization phenomenon, we choose the circuit components $R_{2,3}=1.6\,\mbox{k}\Omega$, $C_{3,5}=10\,\mbox{nF}\pm5\%$, $C_{4,6}=100\,\mbox{nF}\pm5\%$, $L_{1,2}=26\,\mbox{mH}\pm10\%$, and $R_{4,5}=10\,\Omega\pm10\%$. Note that due to the mismatched parameters (components), the two circuits are non-identical. Despite the mismatched parameters, both circuits present chaotic motions when isolated, as depicted in Fig.~\ref{fig3}(c). The two circuits are coupled through the resistor $R_6$, which can be adjusted between $9\,\mbox{k}\Omega$ (strong coupling) and $13\,\mbox{k}\Omega$ (weak coupling) with a high precision ($\sim0.1\Omega$). Still, the currents of the inductors $i_{L_1}$ and $i_{L_2}$ are monitored, respectively, by the voltages $v_{R_4}$ and $v_{R_5}$, and data are acquired by the sampling frequency $f_0=100\,\mbox{kHz}$ for a period of $100\,\mbox{ms}$ in each experiment. 

Setting $R_6=10.2\,\mbox{k}\Omega$, we plot in Fig.~\ref{fig3}(d) the relationship between the voltages $v_{C_3}$ (from circuit 1) and $v_{C_5}$ (from circuit 2). We see that the data are distributed roughly along the diagonal line, indicating that the two circuits are oscillating in a weakly coherent fashion. The synchronization degree of the coupled circuits is evaluated by the time-averaged synchronization error $\delta r=\left<\delta e(t)\right>_T$, with $\delta e=\sqrt{(v_{C_3}-v_{C_5})^2+(v_{C_4}-v_{C_6})^2+(v_{R_4}-v_{R_5})^2}$ the instant synchronization error between the circuits and $\left<\cdot\right>$ the time-average function. For the results shown in Fig.~\ref{fig3}(d), we have $\delta r\approx 0.303 V$. Here the question we are interested in is: {\it given experiments are conducted at only several values of $R_{6}$ and the time series of the sampling states are available, can we anticipate the synchronization degree of the coupled circuits for a random $R_6$ and, furthermore, the variation of the synchronization degree with respect to $R_6$ over a wide range?} The second objective of our present work is to demonstrate that this question can be addressed by the technique of PARC in machine learning.   

\section{Parameter-aware reservoir computing}

\begin{figure}[tbp]
\begin{center}
\includegraphics[width=\linewidth]{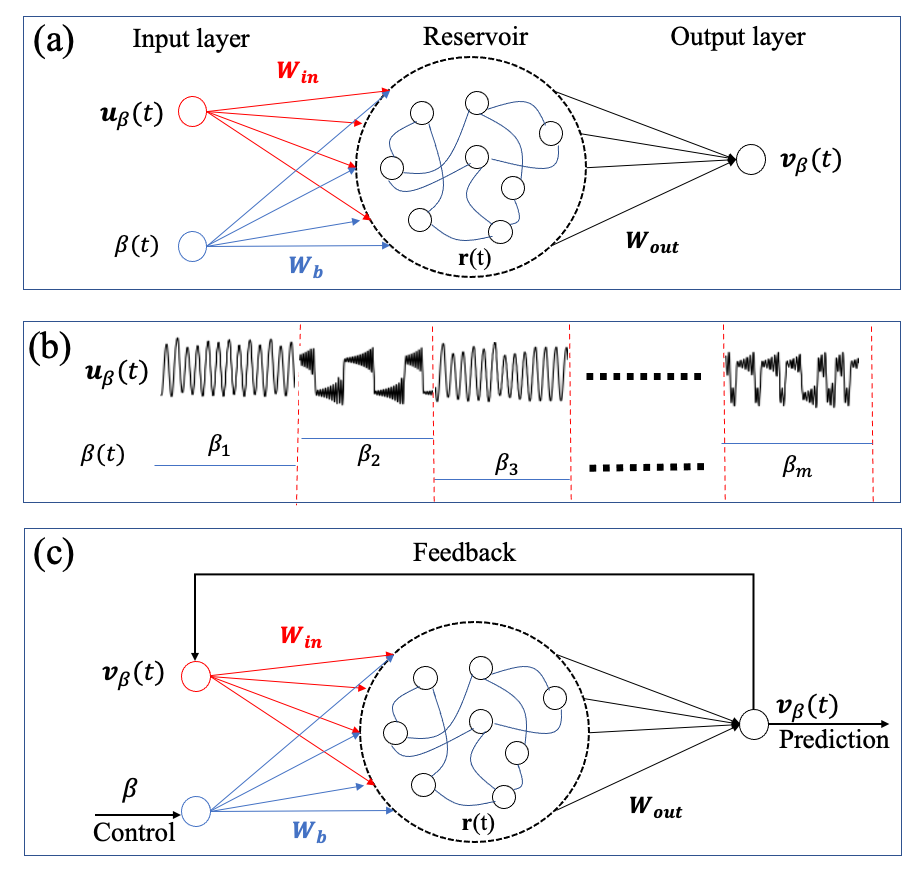}
\centering\caption{Schematic of the PARC technique. (a) The open-loop configuration of the machine in the training phase. (b) Schematic of the training data. (c) The closed-loop configuration of the machine in the predicting phase.} 
\vspace{-0.2cm}
\label{fig4}
\end{center}
\end{figure}

The PARC technique exploited for reconstructing the bifurcation diagrams is generalized from the one proposed in Refs.~\cite{KLW2021,RC:Kim2021,HWFan2021,RC:ZH2021}. Like the conventional RCs, the machine employed here is also constructed by four modules: the $I/R$ layer (input-to-reservoir), the parameter-control channel, the reservoir network, and the $R/O$ layer (reservoir-to-output). The structure of the machine is schematically shown in Fig.~\ref{fig4}(a). The $I/R$ layer is characterized by the matrix $\mathbf{W}_{in}\in\mathbb{R}^{D_r\times D_{in}}$, which couples the input vector $\mathbf{u}_{\beta}(t)\in\mathbb{R}^{D_{in}}$ to the reservoir network. Here, $\mathbf{u}_{\beta}(t)$ denotes the input vector acquired from the target system at time $t$ under the specific bifurcation parameter $\beta$. (For objective one in which the task is to reconstruct the bifurcation diagram of a single circuit, we have $\beta=R$; for objective two in which the task is to anticipate the variation of the synchronization degree of coupled chaotic circuits, we have $\beta=R_6$.) The elements of $\mathbf {W}_{in}$ are randomly drawn from a uniform distribution within the range $[-\sigma, \sigma]$. The parameter-control channel is characterized by the vector $\mathbf{s}=\beta\mathbf{W}_b$, with $\beta$ the control parameter and $\mathbf{W}_b\in \mathbb{R}^{D_{r}}$ the bias vector. The control parameter $\beta$ can be treated as an additional input channel marking the input vector $\mathbf{u}(t)$. The elements of $\mathbf{W}_b$ are also drawn randomly within the range $[-\sigma, \sigma]$. The reservoir network contains $D_r$ nodes, with the initial states of the nodes being randomly chosen from the interval $[-1,1]$. The states of the nodes in the reservoir network, $\mathbf{r}(t)\in \mathbb{R}^{D_r}$, are updated as
\begin{equation}\label{rc1}
\mathbf{r}(t+\Delta t)=(1-\alpha)\mathbf{r}(t)+\alpha\tanh[\mathbf {A}\mathbf{r}(t)+\mathbf{W}_{in}\mathbf{u}_{\beta}(t)+\beta\mathbf{W}_b].
\end{equation}
Here, $\Delta t$ is the time step for updating the reservoir network, $\alpha\in (0,1]$ is the leaking rate, $\mathbf{A}\in \mathbb{R}^{D_r\times D_r}$ is a weighted adjacency matrix representing the coupling relationship between nodes in the reservoir. The adjacency matrix $\mathbf{A}$ is constructed as a sparse random Erd\"{o}s-R\'{e}nyi matrix: with the probability $p$, each element of the matrix is arranged a nonzero value drawn randomly from the interval $[-1,1]$. The matrix $\mathbf{A}$ is rescaled to make its spectral radius equal $\lambda$. The output layer is characterized by the matrix $\mathbf{W}_{out}\in \mathbb{R}^{D_{out}\times D_{r}}$, which generates the output vector, $\mathbf{v}(t)\in \mathbb{R}^{D_{out}}$, according to the equation 
\begin{equation}\label{rc2}
\mathbf{v}(t+\Delta t)=\mathbf{W}_{out}\mathbf{\tilde{r}}(t+\Delta t),
\end{equation}
with $\mathbf{\tilde{r}}\in \mathbb{R}^{D_r}$ the new state vector transformed from the reservoir state (i.e., $\tilde{r}_i=r_i$ for the odd nodes and $\tilde{r}_i=r_i^2$ for the even nodes)~\cite{RC:Pathak2018}, and $\mathbf{W}_{out}$ the output matrix to be estimated by a training process. Except $\mathbf{W}_{out}$, all other parameters of the RC, e.g., $\mathbf{W}_{in}$, $\mathbf{A}$ and $\mathbf{W}_b$, are fixed at the construction. For the sake of simplicity, we set $D_{out}=D_{in}$ in our studies~\cite{RC:Lu2017,RC:Pathak2017,RC:Pathak2018}.

The implementation of PARC consists of three phases: training, validating, and predicting. The mission of the training phase is to find a suitable output matrix $\mathbf{W}_{out}$ so that the output vector $\mathbf{v}(t+\Delta t)$ as calculated by Eq.~(\ref{rc2}) is as close as possible to the input vector $\mathbf{u}(t+\Delta t)$ for $t=(\tau+1)\Delta t,\ldots,(\tau+\hat{L})\Delta t$, with $T_0=\tau\Delta t$ the transient period (used for removing the impact of the initial conditions of the reservoir) and $\hat{L}$ the length of the training series. This is done by minimizing the cost function with respect to $\bm{W}_{out}$~\cite{RC:Lu2017,RC:Pathak2017,RC:Pathak2018}, which gives
\begin{equation}\label{rc3}
\bm{W}_{out}=\bm{U}\bm{V}^T(\bm{V}\bm{V}^T+\eta\mathbb{I})^{-1}.
\end{equation}
Here, $\bm{V}\in \mathbb{R}^{D_{r}\times \hat{L}}$ is the state matrix whose $k$th column is $\bm{\tilde{r}}[(\tau+k)\Delta t]$, $\bm{U}\in \mathbb{R}^{D_{out}\times \hat{L}}$ is a matrix whose $k$th column is $\bm{u}[(\tau+k)\Delta t]$, $\mathbb{I}$ is the identity matrix, and $\eta$ is the ridge regression parameter for avoiding the overfitting.  We note that in the training phase the input data consists of two different time series: (1) the input vector $\mathbf{u}_{\beta}(t)$ representing the state of the target system and (2) the control parameter $\beta(t)$ labeling the condition under which the input vector $\mathbf{u}_{\beta}(t)$ is acquired. In specific, the input vector $\mathbf{u}_{\beta}(t)$ is composed of $m$ segments of length $\hat{n}$, while each segment is a time series obtained from the target system under the specific control parameter $\beta$. As such, the training dataset is a concatenation of the sampling series, and $\beta(t)$ is a step-function of time. The structure of the training data is schematically shown in Fig.~\ref{fig4}(b). 

The machine that performs well on the training data might not perform equally well on the testing data. The finding of the optimal machine performing well on both the training and testing data is the mission for the validating phase. The set of hyperparameters to be optimized in the machine include $D_{r}$ (the size of the reservoir network), $p$ (the density of the adjacency matrix $\mathbf{A}$), $\sigma$ (the range defining the input matrix and the bias vector), $\lambda$ (the spectral radius of the adjacency matrix $\mathbf{A}$), $\eta$ (the regression coefficient), and $\alpha$ (the leaking rate). In our studies, the optimal hyperparameters are obtained by scanning each hyperparameter over a certain range in the parameter space using conventional optimization algorithms such as the Bayesian and surrogate optimization algorithms~\cite{KLW2021}. After finding the optimal machine, we then utilize it to reconstruct the bifurcation diagrams, namely the predicting phase. Shown in Fig.~\ref{fig4}(c) is the flowchart of the machine in the predicting phase. In making the predictions, we replace $\mathbf{u}_{\beta}(t)$ with $\mathbf{v}(t)$ (so that the machine is working in the closed-loop configuration), while setting the control parameter $\beta$ to a specific value of interest. As such, in the predicting phase the machine is still driven by the externally added parameter $\beta$. The output vector $\mathbf{v}(t)$ then gives the predictions, based on which the climate of the system dynamics associated with $\beta$ can be replicated. (Still, before making the predictions, a short transient is discarded to avoid the impact of the initial conditions of the reservoir.) Finally, by tuning $\beta$ in the parameter space, we can reconstruct the whole bifurcation diagram according to the machine predictions.  

\section{Results}

\begin{figure}[tbp]
\begin{center}
\includegraphics[width=0.75\linewidth]{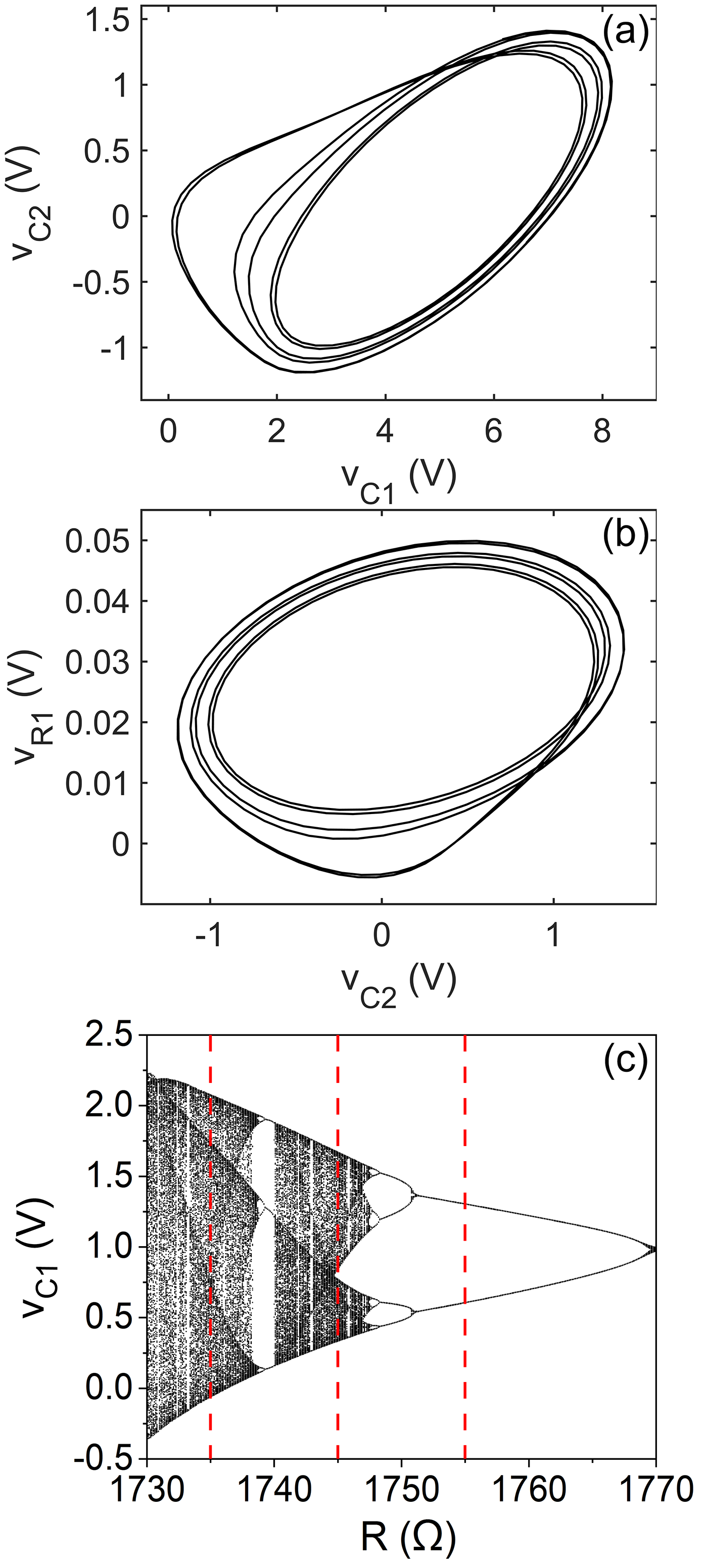}
\centering\caption{Reconstructing the bifurcation diagram of Chua's circuit by the PARC technique. (a,b) The trajectories predicted by the machine for the parameter $R=1.738\,\mbox{k}\Omega$, which is not included in the sampling set. (c) The bifurcation diagram predicted by the PARC technique. Red dashed lines denote the sampling states from which data are measured from experiments.} 
\vspace{-0.2cm}
\label{fig5}
\end{center}
\end{figure}

We first utilize the PARC technique to reconstruct the bifurcation diagram of a single circuit. We begin by choosing the set of sampling states from which the data are acquired from experiments. Previous studies have shown that the performance of PARC is influenced by both the number and the locations of the sampling states~\cite{KLW2021,HWFan2021,RC:ZH2021}. In general, the more the sampling states, the better the machine predictions. Additionally, to replicate the dynamics of a new state that is not included in the sampling set, it is better to choose the sampling states evenly over the parameter space. For demonstration purpose, here we choose $m=3$ sampling states over the bifurcation range plotted in Fig.~\ref{fig5}(c), $R=1.735\,\mbox{k}\Omega$, $1.745\,\mbox{k}\Omega$, and $1.755\,\mbox{k}\Omega$. For each of the sampling states, we record the system evolution for $T=100\,\mbox{ms}$, from which we obtain a time series of $n=5000$ data points. Following the standard strategies in machine learning, we separate the time series into two segments of equal length, with the first half being used as training data and the second half as validating data. The size (length) of the whole training dataset therefore is $\hat{N}=m\times n/2=7500$, so is the validating dataset. (To make the predictions more relevant to the experimental results, here we use the raw data as the input, i.e., the data are not processed.)

We next train the machine and find the optimal set of hyperparameters. In training the machine, the transient series used to remove the impact of the initial conditions of the reservoir contains $\tau=200$ data points (which applies to each of the sampling series in the training data). As such, the total number of data points used for estimating the output matrix $\mathbf{W}_{out}$ is $\hat{L}=m\times \hat{n}=m\times (n/2-\tau)=6900$. To find the optimal set of hyperparameters, we search the hyperparameters over the ranges $D_r\in (200,1000)$, $p\in (0,0.2)$, $\sigma\in (0,1)$, $\lambda\in(0.5,1)$, $\eta\in(1\times 10^{-8},1\times 10^{-2})$, and $\alpha\in(0,1]$ by the Bayesian optimization algorithm. Each set of hyperparameters defines a machine, whose performance is evaluated on the validating data according to the prediction error $\left<|\mathbf{u}(t)-\mathbf{v}(t)|\right>_T$. Still, in evaluating the machine performance by the validating data, a transient series of $\tau=200$ points are used to remove the impact of the initial conditions of the reservoir. For this application, the optimal hyperparameters are $(D_r,p,\sigma,\lambda,\eta,\alpha)=(502,0.15,0.32,0.85,1.2\times 10^{-5},0.54)$, which define the optimal machine to be used for prediction purposes. 

Before employing the trained machine to reconstruct the bifurcation diagram, we check first the capability of the machine in predicting the dynamics of a new state not included in the sampling set. The exampling state we choose is $R=1.738\,\mbox{k}\Omega$. [The trajectories of this state plotted according to experimental data are shown in Figs.~\ref{fig2}(a) and (b).] Setting the control parameter as $\beta=1.738\,\mbox{k}\Omega$, we now operate the machine in the closed-loop configuration [see Fig.~\ref{fig4}(c)]. After a transient period of $\tau=1000$ iterations, the machine begins to output the predictions. The trajectories predicted by the machine are plotted in Figs.~\ref{fig5}(a) and (b). Compared to the smeared trajectories plotted in Figs.~\ref{fig2}(a) and (b), we see in Figs.~\ref{fig5}(a) and (b) that the trajectories show clearly the period-6 orbits. We therefore see that the machine is able to not only infer the dynamics of a new state, but also recover from noise-contaminated signals the true trajectories (i.e., the climate of the system dynamics). We proceed to reconstruct the bifurcation diagram of the circuit by the PARC technique. This is done by increasing the control parameter from $\beta=1.73\,\mbox{k}\Omega$ to $1.77\,\mbox{k}\Omega$ gradually, while for each value of $R$ we collected from the machine output a sequence of $10000$ data points. Shown in Fig.~\ref{fig5}(c) is the bifurcation diagram plotted according to the machine predictions. Compared with the experimentally obtained results [see Fig.~\ref{fig2}(c)], we see that the bifurcation diagram predicted by the machine is of high quality and precision. Specifically, we can infer from the reconstructed bifurcation diagram not only the transition points of the high-order periodic orbits, but also the periodic windows embedded in the chaotic regions.   

We continue to anticipate the synchronization degree of two coupled chaotic Chua circuits by the PARC technique. Still, to generate the training and validating datasets, we acquire from experiments the time series of $m=3$ sampling states, $R_6=9.4\,\mbox{k}\Omega$, $10.2\,\mbox{k}\Omega$ [the state shown in Fig.~\ref{fig3}(d)], and $11\,\mbox{k}\Omega$. Each series contains $n=10000$ data points, with the first half being used as training data and the second half being used as validating data. The transient period of the training phase contains $\tau=500$ data points, and the same transient period is applied in the validating phase. Still, the machine hyperparameters are optimized by the Bayesian optimization algorithm. In this application, the optimal hyperparameters are $(D_r,p,\sigma,\lambda,\eta,\alpha)=(983,4.8\times 10^{-3},0.88,0.39,2.9\times 10^{-3},0.73)$. 

\begin{figure}[tbp]
\begin{center}
\includegraphics[width=0.7\linewidth]{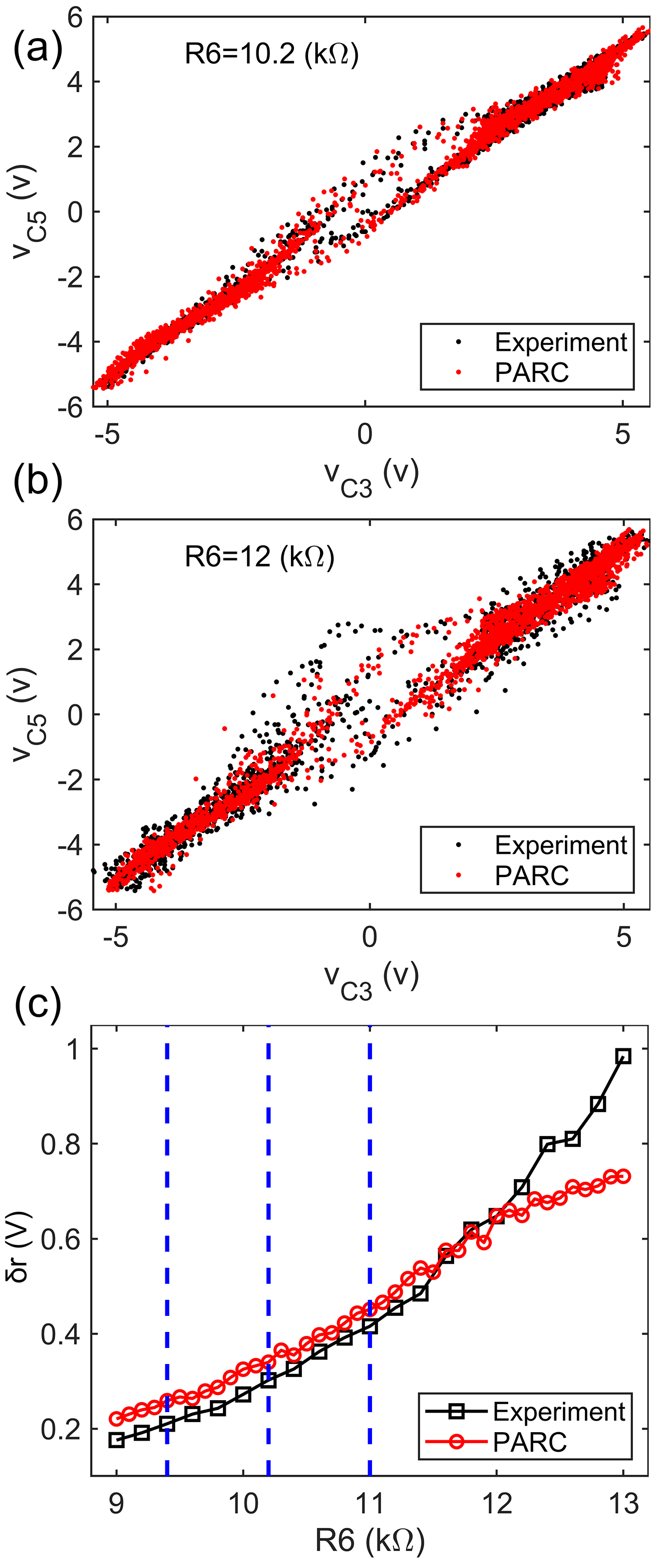}
\centering\caption{Reconstructing the synchronization transition of two coupled chaotic Chua circuits by the PARC technique. The relationship between $v_{C_3}$ and $v_{C_5}$ for (a) $R_6=10.2\,\mbox{k}\Omega$ and (b) $R_6=12\,\mbox{k}\Omega$. Black dots are results acquired from experiments. Red dots are results predicted by the machine. (c) The variation of the synchronization error between the coupled circuits, $\delta r$, with respect to the coupling coefficient, $R_6$. Black squares are results obtained from experiments. Red circles are results predicted by the machine. Blue dashed lines denote the sampling states from which data are measured from experiments.} 
\vspace{-0.4cm}
\label{fig6}
\end{center}
\end{figure}

We check first the capability of the trained machine in replicating the synchronization dynamics of the sampling states. Setting the control parameter as $\beta=10.2\,\mbox{k}\Omega$, we operate the machine in the closed-loop configuration [see Fig.~\ref{fig4}(c)], and estimate from the machine outputs the synchronization error, $\delta r$, between the circuits. The results show that $\delta r\approx 0.34\,V$, which is in good agreement with the experimental results ($\delta r\approx 0.30\,V$). Shown in Fig.~\ref{fig6}(a) is the relationship between $v_{C_3}$ and $v_{C_5}$ for the machine-predicted data (red dots), which is also consistent with the one plotted according to the experimental data (black dots).  

We check next the capability of the machine in anticipating the synchronization climate of a new state not included in the sampling set. To demonstrate, we set $\beta=12\,\mbox{k}\Omega$ and, based on the machine predictions, plot in Fig.~\ref{fig6}(b) the relationship between $v_{C_3}$ and $v_{C_5}$. Compared to the results of $\beta=10.2\,\mbox{k}\Omega$, we see that the synchronization degree between the circuits is clearly decreased for $\beta=12\,\mbox{k}\Omega$. Specifically, for $\beta=12\,\mbox{k}\Omega$, the synchronization error estimated from the machine predictions is $\delta r\approx 0.65\,V$. This estimation is also in good agreement with the experimental result ($\delta r\approx 0.64\,V$), as depicted in Fig.~\ref{fig6}(b).

We finally utilize the machine to anticipate the variation of the synchronization error, $\delta r$, with respect to the coupling coefficient, $R_6$, over a wide range in the parameter space. In doing this, we increase $\beta$ from $9\,\mbox{k}\Omega$ to $13\,\mbox{k}\Omega$ by the increment $\Delta \beta=0.2\,\mbox{k}\Omega$, and for each $\beta$ calculate from the machine outputs the value of $\delta r$. The results are plotted in Fig.~\ref{fig6}(c) (red circles), which shows that with the increase of $\beta$, the value of $\delta r$ is monotonically increased. To validate the predictions, we tune $R_6$ in the experiment over the same range, and for each $R_6$ calculate from the measured data the synchronization error. The experimental results are also plotted in Fig.~\ref{fig6}(c) (black squares). We see that the predicted and experimental results are consistent within the range $R_6\in (9\,\mbox{k}\Omega, 12\,\mbox{k}\Omega)$, but are slightly diverged when $R_6>12\,\mbox{k}\Omega$. The difference between the predicted and experimental results at large $R_6$ is attributed to the large distance between the sampling and testing states, which has been also observed in previous studies~\cite{KLW2021,HWFan2021,RC:ZH2021}.     

\section{Concluding remarks}

In reconstructing the bifurcation diagram of chaotic systems based on measured data, two of the major difficulties encountered in practice are: (1) the signals are contaminated by noise and (2) the signals are acquired at only a few sampling states. The former makes the reconstructed bifurcation diagram coarse and unclear; the latter renders the reconstructed bifurcation diagram fragmented and incomplete. In our present work, by the experimental data of chaotic Chua circuits, we have shown that both difficulties can be well addressed by the technique of PARC proposed recently in machine learning. Two scenarios have been considered and investigated: reconstructing the bifurcation diagram of a single circuit and anticipating the synchronization transition of two coupled chaotic circuits. In the first scenario, we have demonstrated that by the noisy signals acquired at several sampling states, the trained machine is able to reconstruct the whole bifurcation diagram with high precision. The success of the machine in reconstructing the bifurcation diagram is attributed to the noise-filtering effect of the reservoir and the property of transfer learning. Specifically, fed with noisy signals from which the system dynamics can not be inferred directly, the reservoir is able to output a smooth and clear trajectory. And, guided by the parameter-control channel, the knowledge that the machine learned from the time series of the sampling states can be transferred to infer the dynamics of a new state not included in the sampling set. In the second scenario, we have demonstrated that, trained by the noisy signals collected at a handful of coupling parameters, the machine is able to anticipate the variation of the synchronization degree of the coupled circuits with respect to the coupling parameter over a wide range. Whereas the capability of PARC for inferring the dynamics climate of chaotic systems has been well demonstrated in the literature, the previous studies are all based on modeling systems of noise-free signals~\cite{KLW2021,RC:Kim2021,HWFan2021,RC:ZH2021,Roy2022}. Our studies show that this technique can be also applied to noisy signals generated from realistic systems. 

Though our studies demonstrate preliminarily the capability of the PARC technique for reconstructing the bifurcation diagram of realistic chaotic systems, many questions remain to be addressed. First, for convenience and simplicity, we have adopted Chua's circuits as examples to demonstrate the performance of the PARC technique. The applicability of this technique to other real-world chaotic systems is yet to be checked. Second, recent studies show that noise might play a constructive role in the machine learning of chaotic systems~\cite{Noise:AR2021,HWFan2022,Noise:SR2023}. In particular, a stochastic-resonance-like phenomenon has been observed in predicting chaos, where it is shown that the prediction performance can be improved the introducing a certain amount of noise~\cite{Noise:SR2023}. It will be interesting to check whether a similar phenomenon can be observed in the experiments of Chua's circuits. Third, our studies focus on only the low-dimensional chaotic systems (a single Chua's circuit and two coupled chaotic Chua circuits). It remains not clear whether the same PARC technique can be applied to high-dimensional chaotic systems, e.g., spatially extended chaotic systems and large-size complex networks of coupled oscillators. In applying the technique to high-dimensional chaotic systems, one difficulty concerns the super size of the reservoir network. One possible approach to addressing this difficulty could be adopting the scheme of parallel RC~\cite{RC:Pathak2018}, which, however, might need a significant modification of the machine structure. Finally, an important feature of many real-world chaotic systems is that their asymptotic dynamics are dependent on the initial conditions, namely the property of multistability~\cite{Multistable}. The application of the PARC technique to reconstruct the bifurcation diagrams of multistable chaotic systems, probably by incorporating some additional modules to the current machine, is another interesting topic warranting further studies. 
\vspace{-0.3cm}

\begin{acknowledgments}
This work was supported by the National Natural Science Foundation of China (NNSFC) under Grant Nos.~12275165 and 12105165. XGW was also supported by the Fundamental Research Funds for the Central Universities under Grant No.~GK202202003.
\end{acknowledgments}





\begin{thebibliography}{99}
\bibitem{Ott:Book}E. Ott, {\it Chaos in Dynamical Systems} (Cambridge Universitiy Press, Combridge, 2002).

\bibitem{MS:Book}M. Scheffer, {\it Critical Transitions in Nature and Society} (Princeton University Press, 2009).


\bibitem{TP:MG2006}M. Gladwell, {\it The Tipping Point: How Little Things Can Make a Big Difference} (Little, Brown, 2006).

\bibitem{TP:Review2009}M. Scheffer, J. Bascompte, W. A. Brock, V. Brovkin, S. R. Carpenter, V. Dakos, H. Held, E. H. van Nes, M. Rietkerk, and G. Sugihara, Early-warning signals for critical transitions, Nature {\bf 461}, 53 (2009).

\bibitem{TP:Climate1}T. M. Lenton, H. Held, E. Kriegler, J. W. Hall, W. Lucht, S. Rahmstorf, and H. J. Schellnhuber, Tipping elements in the Earth’s climate system, Proc. Natl. Acad. Sci. USA  {\bf 105}, 1786 (2008).

\bibitem{TP:Climate2}T. M. Lenton, J. Rockström, O. Gaffney, S. Rahmstorf, K. Richardson, W. Steffen, and H. J. Schellnhuber, Climate tipping points — too risky to bet against, Nature {\bf 575}, 592 (2019).

\bibitem{TP:Eco1}M. Scheffer, S. Carpenter, J. A. Foley, C. Folke, and B. Walker, Catastrophic shifts in ecosystems, Nature {\bf 413}, 591 (2001).

\bibitem{TP:Eco2}M. Hirota, M. Holmgren, E. H. V. Nes, and M. Scheffer, Global resilience of tropical forest and savanna to critical transitions, Science {\bf 334}, 232 (2011).

\bibitem{TP:FinancialSys1}A. G. Haldane and R. M. May, Systemic risk in banking ecosystems, Nature {\bf 469}, 351 (2011).

\bibitem{TP:FinancialSys2}A. Majdandzic, L. a. Braunstein, C. Curme, I. Vodenska, S. Levy-Carciente, H. E. Stanley, and S. Havlin, Multiple tipping points and optimal repairing in interacting networks, Nature Commun. {\bf 7}, 10850 (2016).


\bibitem{BD:FD1980}F. Takens, in {\it Detecting Strange Attractors in Turbulence: Dynamical Systems and Turbulence, Warwick}, edited by D. A. Rand, and L. S. Young (Springer, Berlin, 1980).

\bibitem{BD:NHP1980}N. H. Packard, J. P. Crutchfield, J. D. Farmer, and R. S. Shaw, Geometry from a time series, Phys. Rev. Lett. {\bf 45}, 712 (1980).

\bibitem{BD:RT1994}R. Tokunaga, S. Kajiwara, and T. Matsumoto, Reconstructing bifurcation diagrams only from time-waveforms, Physica D {\bf 79}, 348 (1994).


\bibitem{BD:EB1999}E. Bagarinao, K. Pakdaman, T. Nomura, and S. Sato, Reconstructing bifurcation diagrams from noisy time series using nonlinear autoregressive models, Phys. Rev. E {\bf 60}, 1073 (1999).

\bibitem{BD:GL2004}G. Langer and U. Parlitz, Modeling parameter dependence from time series, Phys. Rev. E {\bf 70}, 056217 (2004).

\bibitem{BD:RC2019}R. Cestnik and M. Abel, Inferring the dynamics of oscillatory systems using recurrent neural networks, Chaos {\bf 29}, 063128 (2019).

\bibitem{RC:Follmann2019}R. Follmann and R. Epaminondas, Predicting slow and fast neuronal dynamics with machine learning, Chaos {\bf 29}, 113119 (2019).

\bibitem{BD:YI2020}Y. Itoh, S. Uenohara, M. Adachi, T. Morie, and K. Aihara, Reconstructing bifurcation diagrams only from time-series data generated by electronic circuits in discrete-time dynamical sys- tems, Chaos {\bf 30}, 013128 (2020).

\bibitem{RC:ZHScienceChina2021}H. Zhao, Inferring the dynamics of ``black-box" systems using a learning machine, Sci. China-Phys. Mech. Astron. {\bf 64}, 270511 (2021).

\bibitem{KLW2021}L.-W. Kong, H. Fan, C. Grebogi, and Y.-C. Lai, Machine learning prediction of critical transition and system collapse, Phys. Rev. Res. \textbf{3}, 013090 (2021).

\bibitem{RC:Kim2021}J. Z. Kim, Z. Lu, E. Nozari, G. J. Pappas, and D. S. Bassett, Teaching recurrent neural networks to infer global temporal structure from local examples, Nat. Mach. Intell. {\bf 3}, 316 (2021).

\bibitem{HWFan2021}H. Fan, L.-W. Kong, Y.-C. Lai, and X. G. Wang, Anticipating synchronization with machine learning, Phys. Rev. Res. \textbf{3}, 023237 (2021).

\bibitem{RC:ZH2021}H. Zhang, H. Fan, L. Wang, and X. G. Wang, Learning Hamiltonian dynamics with reservoir computing, Phys. Rev. E {\bf 104}, 024205 (2021).

\bibitem{Roy2022}M. Roy, S. Mandal, C. Hens, A. Prasad, N. V. Kuznetsov, and M. D. Shrimali, Model-free prediction of multistability using echo state network, Chaos {\bf 32}, 101104 (2022).


\bibitem{Reverse:SLB2016}S. L. Brunton, J. L. Proctor, and J. N. Kutz, Discovering governing equations from
data: Sparse identification of nonlinear dynamical systems, Proc. Natl. Acad. Sci. USA {\bf 113}, 3932 (2016).

\bibitem{CS:WW2016}W. Wang, Y.-C. Lai, and C. Grebogi, Data based identification and prediction of nonlinear and complex dynamical systems, Phys. Rep. {\bf 644}, 1 (2016).

\bibitem{GHU:2018}T. Y. Chen, Y. Chen, H. J. Yang, J. H. Xiao, and G. Hu, Reconstruction of dynamic structures of experimental setups based on measurable experimental data only, Chinese Physics B {\bf 27}, 030503 (2018).


\bibitem{RC:Maass2002}W. Maass, T. Natschlager, and H. Markram, Real-time computing without stable states: A new framework for neural computation based on perturbations, Neural Comput. {\bf 14}, 2531 (2002).

\bibitem{RC:Jaeger}H. Jaeger and H. Haas, Harnessing nonlinearity: Predicting chaotic systems and saving energy in wireless communication, Science {\bf 304}, 78 (2004).

\bibitem{RC:Lu2017}Z. Lu, J. Pathak, B. Hunt, M. Girvan, R. Brockett, and E. Ott, Reservoir observers: Model-free inference of unmeasured variables in chaotic systems, Chaos {\bf 27}, 041102 (2017).

\bibitem{RC:Pathak2017}J. Pathak, Z. Lu, B. Hunt, M. Girvan, and E. Ott, Using machine learning to replicate chaotic attractors and calculate Lyapunov exponents from data, Chaos {\bf 27}, 121102 (2017).

\bibitem{RC:Pathak2018}J. Pathak, B. Hunt, M. Girvan, Z. Lu, and E. Ott, Model-free prediction of large spatiotemporally chaotic systems from data: A reservoir computing approach, Phys. Rev. Lett. {\bf 120}, 024102 (2018).

\bibitem{RC:SynSmall2019}T. Weng, H. Yang, C. Gu, J. Zhang, and M. Small, Synchronization of chaotic systems and their machine-learning models, Phys. Rev. E {\bf 99}, 042203 (2019).

\bibitem{RC:Fan2020}H. Fan, J. Jiang, C. Zhang, X. G. Wang, and Y.-C. Lai, Long-term prediction of chaotic systems with machine learning, Phys. Rev. Res. {\bf 2}, 012080(R) (2020).

\bibitem{RC:Book}K. Nakajima and I. Fischer, {\it Reservoir Computing: Theory, Physical Implementations, and Applications} (Springer, Singapore, 2021).

%
%

\bibitem{NoiseFilter:2020}N. A. K. Doan, W. Polifke, and L. Magri, Physics-informed echo state networks, J. Comput. Sci. {\bf 47}, 101237 (2020).


\bibitem{NoisePDC:2021}D. Patel, D. Canaday, M. Girvan, A. Pomerance, and E. Ott, Using machine learning to predict statistical properties of non-stationary dynamical processes: system climate, regime transitions, and the effect of stochasticity, Chaos {\bf 31}, 033149 (2021).

\bibitem{Noise:Nathe2023}C. Nathe, C. Pappu, N. A. Mecholsky, J. D. Hart, T. Carroll, and F. Sorrentino, Reservoir Computing with Noise, Chaos {\bf 33}, 041101 (2023).

\bibitem{Noised:Lin2022}Z. Lin, Y. Liang, J. Zhao, J. Li, and T. Kapitaniak, Prediction of dynamic systems driven by Lévy noise based on deep learning, Nonlinear Dyn. {\bf 111}, 1511 (2023).

\bibitem{Noise:CB1995}C. Bishop, Training with noise is equivalent to Tikhonov regularization, Neu. Comp. {\bf 7}, 108 (1995).

\bibitem{Noise:AR2021}A. Röhm, D. J. Gauthier, and I. Fischer, Model-free inference of unseen attractors: Reconstructing phase space features from a single noisy trajectory using reservoir computing, Chaos {\bf 31}, 103127 (2021).

\bibitem{HWFan2022}H. Fan, L. Wang, Y. Du, Y. F. Wang, J. H. Xiao, and X. G. Wang, Learning the dynamics of coupled oscillators from transients, Phys. Rev. Res. \textbf{4}, 013137 (2022).

\bibitem{Noise:SR2023}Z.-M. Zhai, L.-W. Kong and Y.-C. Lai, Emergence of a stochastic resonance in machine learning, Phys. Rev. Res. {\bf 5}, 033127 (2023).

\bibitem{ChuaModel}L. O. Chua, Chua’s circuit: an overview ten years later, J. Circuits Syst. Comput. {\bf 4}, 117 (1994).

\bibitem{ChuaExp1}M. P. Kennedy, Robust OP Amp realization of Chua's circuit, Frequenz {\bf 46}, 66 (1992).

\bibitem{ChuaExp2}L. O. Chua, L. Kocarev, K. Eckart, and M. Itoh, Experimental chaos synchronization in Chua’s circuit, Int. J. Bif. Chaos {\bf 2}, 705 (1992).

\bibitem{ChuaExp3}M. S. Baptista, T. P. Silva, J. C. Sartorelli, and I. L. Caldas, Phase synchronization in the perturbed Chua circuit, Phys. Rev. E {\bf 67}, 056212 (2003).

\bibitem{Multistable}A. N. Pisarchik and U. Feudel, Control of multistability, Phys. Rep. {\bf 540}, 167 (2014).

\end{thebibliography}
\end{document}